\documentclass[prl,twocolumn,superscriptaddress,showpacs]{revtex4}

\usepackage{amsmath,amssymb,graphicx,color,bm,soul}
\usepackage[usenames,dvipsnames]{xcolor}

\begin{document}

\author{O. Kyriienko}
\affiliation{Niels Bohr Institute, University of Copenhagen, Blegdamsvej 17,DK-2100 Copenhagen, Denmark}

\author{T. C. H. Liew}
\affiliation{Division of Physics and Applied Physics, Nanyang Technological University 637371, Singapore}

\title{Exciton-Polariton Quantum Gates Based on Continuous Variables}

\begin{abstract}
We propose a continuous variable analog of quantum controlled-NOT gates based on a system of exciton-polaritons in semiconductor microcavities. This can be realized by the engineering of parametric interaction between control and target polariton modes, which can be varied in time. As an explicit setup we use a system of dipolaritons, which allows for enhancement of parametric interaction by auxiliary classical fields and scalable multigate system realization. The calculated fidelity is shown to exceed $99\%$ for realistic system parameters.
\end{abstract}



\maketitle

{\it Introduction.---}Quantum controlled NOT (CNOT) logic gates are universal elements in quantum computation, in principle allowing the implementation of any quantum algorithm (when supplemented with single qubit rotations). Their construction in physical systems is challenging since one requires a system with both limited dephasing and strong nonlinearity. Early realizations of CNOT gates made use of nuclear magnetic resonance in molecules~\cite{Vandersypen2001} or post-selection in linear optical systems~\cite{OBrien2003}, which were later reduced in size on photonic chips~\cite{Politi2009}. More recently, demonstrations of quantum gates and small quantum circuits were achieved using a variety of systems, including ion traps~\cite{Benhelm2008,Blatt2008,Monroe2013,Harty2014}, nitrogen vacancy centers~\cite{vanderSar2012}, and superconducting qubits~\cite{DiCarlo2009,Federov2012,Steffen2013,Barends2014}.

Semiconductor systems have long been valued in information processing for their compact sizes, which are particularly important when one aims at building circuits with large numbers of gates. In particular, semiconductor microcavities containing quantum wells are quasi-two-dimensional structures with micron thicknesses. These structures also offer the opportunity to hybridize the properties of photons and excitons, generating new exciton-polariton quasiparticles with decay time exceeding tens of picoseconds. The presence of an excitonic component facilitates nonlinear interaction between polaritons, which has led to experimental observation of quantum optical effects such as squeezing~\cite{Karr2004} and non-classical correlations~\cite{Savasta2005} for a coherently driven polariton system. Additionally, macroscopically populated polaritonic modes were suggested to mimic the two-level qubit system~\cite{Demirchyan2014} and analog CNOT gate~\cite{Solnyshkov2014}. At the same time, currently existing polaritonic samples do not possess the strong single polariton nonlinearity required for the conventional blockade mechanism~\cite{Verger2006}, and most quantum effects at the weak single polariton nonlinearity level need to exploit interference effects for realization of an unconventional polariton blockade~\cite{Liew2010,Bamba2011} or generation of entangled states~\cite{Liew2013}.

The aim of this paper is to introduce an alternative route towards construction of polariton quantum CNOT gates. To avoid the need of single particle control and detection, we choose to encode information in the continuous amplitude and phase variables of polariton fields. First, a continuous variable quantum CNOT gate operator is realized by engineering a two-mode parametric interaction Hamiltonian, which can be controlled in time. Second, it is well-known that it is difficult to observe nonlinear effects when small numbers of polaritons are involved. For this reason we exploit a mechanism of nonlinear enhancement that uses a macroscopic density to amplify scattering processes, while quantum information is maintained in separated low density quantum modes. The proposal allows to achieve CNOT operation fidelity as high as $99\%$ for realistic system parameters.

Finally, we consider the scalability of our scheme, where multiple quantum gates can be cascaded one after another, in principle allowing the construction of arbitrary algorithms. Typically one imagines a quantum circuit as a network of quantum logic gates separated in space~\cite{Kimble2008}. This requires that signals travel spatially between distant nodes of the network. However, spatially propagating polaritons would experience losses as they scatter with disorder and experience dispersion. While superfluidity~\cite{Amo2009,Amo2009b,Sanvitto2010} and soliton wavepackets~\cite{Sich2011} have been shown to overcome these effects in microcavities, they imply a macroscopic classical polariton state, which cannot itself encode quantum information. In the present proposal we use a reciprocal space encoding of quantum nodes, negating the need for any spatial propagation. Polariton modes are distinguished by different momenta and logic gates exploit momentum conservation rules to connect particular modes. Rather than being physically fixed, the gates are enacted by the application of a known pulse sequence, which could be controlled by a spatial light modulator. This brings the additional feature of being able to reconfigure the quantum circuit.

{\it Definitions.---}The theory of quantum information with continuous variables~\cite{Braunstein2005} is well-developed and uses the same fundamental features of quantum computation, namely superposition and entanglement, to achieve similar aims to qubit based methods. Working with continuous variables, the analogous definition of the quantum CNOT gate is given by the operator $\hat{\mathrm{CN}}=e^{-i\hat{q}_1\hat{p}_2}$ \cite{Braunstein1998}, which acts simultaneously on two quantum fields, $\hat{a}_1$ and $\hat{a}_2$, where the amplitude and phase operators are defined by $\hat{q}_n=\left(\hat{a}_n+\hat{a}_n^\dagger\right)/\sqrt{2}$ and $\hat{p}_n=-i\left(\hat{a}_n-\hat{a}_n^\dagger\right)/\sqrt{2}$. The CNOT gate can also be defined by its action on amplitude eigenstates $\hat{\mathrm{CN}}\left|q_1,q_2\right>=\left|q_1,q_2+q_1\right>$, demonstrating analogy with CNOT gates: the first quantum field $\hat{a}_1$ acts as a control field, which adds to the amplitude of the target field $\hat{a}_2$.

{\it Single gate scheme.---} To realize a single CNOT gate, let us consider the Hamiltonian:
\begin{equation}
\hat{\mathcal{H}}=\alpha\left(\psi^{*2}\hat{a}_1\hat{a}_2+\psi^2\hat{a}^\dagger_1\hat{a}^\dagger_2\right)-J\left(\hat{a}_1^\dagger\hat{a}_2+\hat{a}_2^\dagger\hat{a}_1\right).\label{eq:Ham1}
\end{equation}
It represents a parametric scattering from macroscopically occupied modes $\psi$ into two quantum modes $\hat{a}_1$ and $\hat{a}_2$, which have an additional linear coupling between them. If we imagine that $\hat{a}_1$ and $\hat{a}_2$ represent polariton modes with equal and opposite wavevectors, then the first term in the above Hamiltonian can be arranged making use of interbranch scattering schemes such as those studied theoretically in standard planar microcavities~\cite{Ciuti2004} and parametrically driven cavities~\cite{Bardyn2012}, or experimentally in triple microcavities~\cite{Diederichs2006}. In such cases the field $\psi$ would be a polariton field with zero in-plane wavevector, and $\alpha$ is a constant parameterizing the strength of polariton-polariton interactions. We will assume that $\psi$ can be controlled optically, and ideally the linear coupling term $J$ could also be controlled. Both resonant~\cite{Amo2010} and non-resonant~\cite{Wertz2010} excitation have been used to optically generate polariton potentials, leading to the optical control of polaritons in real-space~\cite{Tosi2012,Gao2012,Schmutzler2014}, and creating the desired linear coupling $J$ between the modes $\hat{a}_1$ and $\hat{a}_2$.

Since the amplitude and phase of $\psi$ can be tuned, let us consider the case $\alpha\psi^2=-J$. The phase reference of the second mode can be changed $\hat{a}_2\mapsto i\hat{a}_2$ following the phase of an input. Then, the Hamiltonian becomes,
\begin{equation}
\hat{\mathcal{H}}=J\left(-i\hat{a}_1\hat{a}_2+i\hat{a}^\dagger_1\hat{a}^\dagger_2-i\hat{a}_1^\dagger\hat{a}_2+i\hat{a}_2^\dagger\hat{a}_1\right)=2J\hat{q}_1\hat{p}_2,
\label{eq:HamiltonianCNOT}
\end{equation}
and the corresponding evolution operator reads $\hat{\mathcal{U}}=e^{-i\hat{\mathcal{H}}t/\hbar}=e^{-i2J\hat{q}_1\hat{p}_2t/\hbar}$. By switching on and off the couplings the evolution time could be set to $t\rightarrow \tau_0=\hbar/2J$, in which case the unitary evolution operator corresponds to the quantum CNOT gate. The result of CNOT gate operation can then be monitored by measuring the amplitude of a target field~\cite{SM}.
We note that while the interaction energy between a pair of polaritons $2\alpha$ may be limited, it is the quantity $\alpha\psi^2$ that determines the relevant coupling strength~\cite{Kyriienko2014}, where weak interaction is effectively amplified by the macroscopic field $\psi$.

{\it Gate fidelity.---}In the ideal case of unitary evolution with precisely timed parametric interaction, which we described before, the polaritonic system can work as a perfect continuous-variable CNOT gate. However, in real systems we identify the main mechanisms responsible for imperfect gate operation to be the decay of polaritons and the imprecision of the interaction constant control.

In the presence of decoherence, the evolution of any expectation value is given by:
\begin{align}
\notag
i\hbar\frac{d\langle\hat{A}\rangle}{dt}=&\left<\left[\hat{A},\hat{\mathcal{H}}\right]+\frac{i\Gamma}{2}\sum_n\mathcal{L}_{\hat{a}_n}[\hat{A}]+\frac{iP}{2}\sum_n\mathcal{L}_{\hat{a}^\dagger_n}[\hat{A}] \right. \\ &\left.+\frac{i\Gamma_{P}}{2}\sum_n\mathcal{L}_{\hat{a}_n^\dagger \hat{a}_n}[\hat{A}]\right>,
\label{eq:Heisenberg}
\end{align}
where $\Gamma$ denotes a dissipation rate of polaritonic modes, and the Lindblad superoperator is defined as $\mathcal{L}_{\hat{a}}[\hat{A}]=2\hat{a}^\dagger\hat{A}\hat{a}-\hat{a}^\dagger\hat{a}\hat{A}-\hat{A}\hat{a}^\dagger\hat{a}$, and Lindbladian $\mathcal{L}_{\hat{a}^\dagger\hat{a}}[\hat{A}]=2\hat{a}^\dagger \hat{a} \hat{A} \hat{a}^\dagger \hat{a}-\hat{a}^\dagger\hat{a} \hat{a}^\dagger\hat{a} \hat{A}-\hat{A}\hat{a}^\dagger\hat{a} \hat{a}^\dagger\hat{a}$ corresponds to a pure dephasing with rate $\Gamma_{P}$. For the sake of generality, we also introduced an additional \emph{incoherent} pumping at rate $P$ with conjugate Lindbladian $\mathcal{L}_{\hat{a}^\dagger}[\hat{A}]$, which is responsible for an incoherent replenishing of polaritonic states, and may arise from a presence of a thermal reservoir.

Using Eq.~(\ref{eq:Heisenberg}), we derive a closed set of evolution equations for the amplitude and phase expectation values:
\begin{align}
\label{eq:dq1dt}
i\hbar\frac{d\langle \hat{q}_1\rangle}{dt}&=\frac{i\left(P-\Gamma-\Gamma_{P}\right)}{2}\langle \hat{q}_1\rangle,\\
\label{eq:dq2dt}
i\hbar\frac{d\langle \hat{q}_2\rangle}{dt}&=2iJ\langle\hat{q}_1\rangle+\frac{i\left(P-\Gamma-\Gamma_{P}\right)}{2}\langle \hat{q}_2\rangle,\\
\label{eq:dp1dt}
i\hbar\frac{d\langle \hat{p}_1\rangle}{dt}&=-2iJ\langle\hat{p}_2\rangle+\frac{i\left(P-\Gamma-\Gamma_{P}\right)}{2}\langle \hat{p}_1\rangle,\\
\label{eq:dp2dt}
i\hbar\frac{d\langle \hat{p}_2\rangle}{dt}&=\frac{i\left(P-\Gamma-\Gamma_{P}\right)}{2}\langle \hat{p}_2\rangle,
\end{align}
which are readily solved analytically. One sees that for $P=\Gamma$ the mean-field amplitudes of the polaritons no longer decay, while in the regime $P>\Gamma$ one should account also for nonlinear losses~\cite{Keeling2008} to prevent the amplitudes growing indefinitely. Furthermore, one can write a closed set of evolution equations for the second order correlators ($\langle \hat{q}_1^2\rangle$, $\langle \hat{q}_1\hat{q}_2\rangle$, $\langle \hat{q}_2^2\rangle$, $\langle \hat{p}_1^2\rangle$, etc.). These equations are given in the Supplemental Material \cite{SM}, together with their analytical solution.

To assess the performance of the quantum gate, we consider a set of displaced squeezed vacuum states as inputs $\boldsymbol{\rho}_\mathrm{in}=\left|\psi_1\right>\left<\psi_1\right|\otimes\left|\psi_2\right>\left<\psi_2\right|$, with $\left|\psi_n\right>=\hat{D}(q^\mathrm{in}_n,p^\mathrm{in}_n)\hat{S}(r)\left|0\right>$, where $\hat{D}(q^\mathrm{in}_n,p^\mathrm{in}_n)=\exp\left(\langle a^\mathrm{in}_n\rangle\hat{a}_n^\dagger-\langle a^\mathrm{in}_n\rangle^*\hat{a}_n\right)$ is the displacement operator providing the amplitude $q^\mathrm{in}_n$ and phase $p^\mathrm{in}_n$ of the input mean-fields $\langle a^\mathrm{in}_n\rangle$. $\hat{S}(r)=\exp\left(\frac{r}{2}\hat{a}^{\dagger2}-\frac{r}{2}\hat{a}^2\right)$ is the squeezing operator with squeezing parameter $r$, and $\left|0\right>$ is the vacuum state. The ideal output state is $\boldsymbol{\rho}_\mathrm{ideal}=\left|q^\mathrm{in}_1,q^\mathrm{in}_1+q^\mathrm{in}_2\right>\left<q^\mathrm{in}_1,q^\mathrm{in}_1+q^\mathrm{in}_2\right|$. The actual state $\boldsymbol{\rho}(t)$ obtained in the presence of dissipation and incoherent pumping is characterized by its fidelity with $\boldsymbol{\rho}_\mathrm{ideal}$:
\begin{equation}
F\left(\boldsymbol{\rho}_\mathrm{ideal},\boldsymbol{\rho}(t)\right)=\mathrm{Tr}\left[\sqrt{\sqrt{\boldsymbol{\rho}_\mathrm{ideal}}\boldsymbol{\rho}(t)\sqrt{\boldsymbol{\rho}_\mathrm{ideal}}}\right].
\end{equation}

The fidelity can be calculated from the covariance matrix~\cite{Marian2012,Spedalieri2013}, which is directly obtained by knowing all the second order correlators.
\begin{figure}
\includegraphics[width=\linewidth]{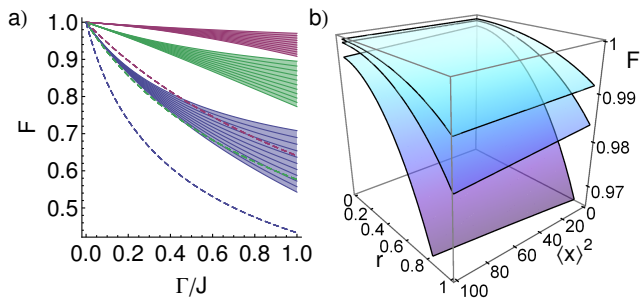}
\caption{(color online) Fidelity of a quantum CNOT gate for increasing dissipation rate. The different colors correspond to different amounts of input squeezing (magenta: r=0, green: r=0.5, and blue: r=1). The solid curves show the range of fidelities for total intensity increasing from $\langle x\rangle^2=0$ to $\langle x\rangle^2=10$, where the highest fidelity is obtained for the squeezed vacuum input. Dashed curves show the same result in the presence of an additional incoherent pumping, chosen with rate $P=\Gamma$. b) Dependence of the minimum fidelity of the input squeezing $r$ and total intensity $\langle x\rangle^2$. The different surfaces correspond to different values of $\Gamma/J=(0.01,0.02,0.05)$. Smaller values of $\Gamma/J$ leading to very high fidelities are expected in samples with high quality factor.}
\label{fig:Fidelity}
\end{figure}
The fidelity of the polaritonic CNOT gate as a function of decay rate is shown in Fig.~\ref{fig:Fidelity}(a), for different values of the squeezing parameter $r$. For each value of $r$ we minimize the fidelity over input states with a fixed maximum total intensity $\langle x\rangle^2=\langle q_1\rangle^2+\langle p_1\rangle^2+\langle q_2\rangle^2+\langle p_2\rangle^2$. We assume the optimum operation time $\tau_0=\hbar/2J$. The fidelity is seen to drop from unity (corresponding to a perfect CNOT gate) monotonically as the ratio of the decay rate to the coupling strength $\Gamma/J$ is increased, or when input states with a higher squeezing or total intensity are considered, as illustrated in Fig.~\ref{fig:Fidelity}(b).

The decrease of the fidelity when operating with more highly squeezed states is very natural. In phase space, the Wigner function of the unsqueezed state is a circular Gaussian, while the infinitely squeezed state is a thin line. The decay has the effect of smearing any squeezed state into the unsqueezed state with $\langle q_n^2\rangle=\langle p_n^2\rangle=1/2$. In fact the state is driven toward the unsqueezed state at a rate proportional to $1/2-\langle q^2 \rangle$ (as can be seen from writing the correlator evolution explicitly), that is, highly squeezed states are most quickly deformed in phase space.
\begin{figure}
\includegraphics[width=\linewidth]{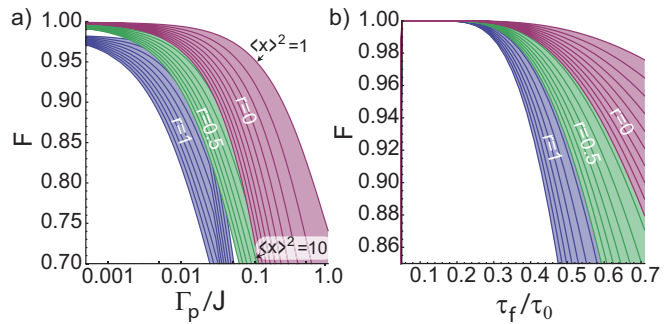}
\caption{(color online) Range of CNOT fidelities calculated for input states of total intensity increasing from $\langle x\rangle^2=1$ to $\langle x\rangle^2=10$, accounting for a pure dephasing and non-ideal gate pulse. a) Fidelity as a function of pure dephsing rate for input states with squeezing parameter $r = 0, 0.5, 1$. b) Gate fidelity in square $J(t)$ pulse operation mode plotted as a function of signal edge time to optimal operation time ratio, $\tau_f/t_0$. }
\label{fig:Fidelity_2}
\end{figure}

The dashed curves show the fidelities calculated with a non-zero incoherent pump, which attempts to compensate the losses in the system ($P=\Gamma$). Interestingly, in this case the fidelity no longer depends on the intensity of the initial state, however it becomes worse due to the incoherent pump. While the incoherent pump can compensate fully the loss of mean-field amplitudes, it can not compensate the loss of quantum correlations caused by dissipation. Allowing for different values of $P$, one finds that the optimum fidelity appears for $P=0$.

Next, we plot the dependence of gate fidelity on the pure dephasing rate. The results are shown in Fig.~\ref{fig:Fidelity_2}(a) assuming the decay rate $\Gamma/J = 0.02$, input states with $r=0,0.5,1$, and varying total intensity ranging from $\langle x\rangle^2=1$ to $\langle x\rangle^2=10$. These suggest that a low level of pure dephasing $\Gamma_{P}/J < 0.05$ is required for high quality CNOT operation.

Additionally, we study the influence of imperfect timing of the gate pulse on the fidelity of the continuous variable CNOT gate. We start considering a square pulse with finite pulse edges described by $J(t)=J_0 f(t)[1-f(t-\tau_0)]$, where $f(t)=[\exp(-t/\tau_f)+1]^{-1}$, with $\tau_f$ being pulse edge time, and $J_0$ denotes time independent interaction constant. The fidelity as a function of pulse edge time is shown in Fig.~\ref{fig:Fidelity_2}(b) for an ideal case of negligible losses. The results suggest that an accurate control of the gate pulse is required at a time scale below $\hbar/J_0$, where pulses with sharp edges, $\tau_f/\tau_0 < 0.3$ do not contribute to degradation of the gate fidelity. 

Finally, we discuss the experimental feasibility of the proposed scheme, with the main parameters being the polariton lifetime, nonlinear coupling constant, pure dephasing, and characteristic switching time. Polariton lifetimes $\tau_{\mathrm{dec}}$ in the range of hundreds of picoseconds have been reported recently~\cite{Nelson2013,Steger2014}, with the corresponding decay rate being in the $\mu$eV range. When controlled with a classical field, nonlinearities can be tuned to be in the sub-meV range.
Taking $J=0.2$~meV and $\tau_{\mathrm{dec}}=65$~ps gives the ratio of $\Gamma/J \approx 0.02$, and sets the characteristic switching time to a picosecond range, which shall be achievable with an optical control of the coupling constant. The pure dephasing in polaritonic system was estimated to be at $\Gamma_P = 0.2~\mu$eV level \cite{Savona1997}, yielding $\Gamma_{P}/J = 10^{-3}$. This makes high fidelities of $99\%$ feasible for relevant input states of $\langle x^2 \rangle < 10$ and $r \approx 0.5$.

\begin{figure}
\includegraphics[width=0.6\linewidth]{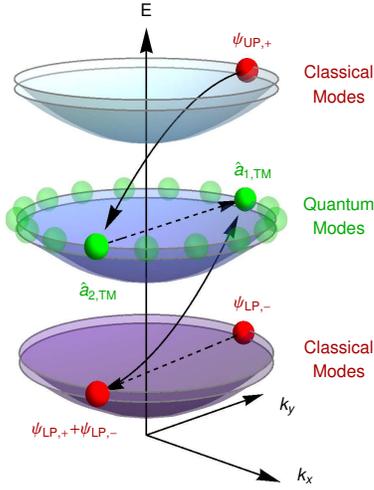}
\caption{(color online) Scheme for a CNOT gate implemented between a selected pair of modes. The plot shows 2D dispersion of three polaritonic modes, each split in two branches for different polarization components.}
\label{fig:ScalableScheme}
\end{figure}

{\it Multi-gate system: dipolariton setup.---}We now discuss a scalable scheme of CNOT gates, which requires a set of multiple quantum modes and the ability to apply successive gates between chosen pairs of modes. Toward this aim let us consider a dipolariton system~\cite{Crisofolini2012}, in which two types of exciton modes (direct and indirect) are coupled to a cavity mode in a microcavity resulting in three dispersion branches as illustrated in Fig.~\ref{fig:ScalableScheme}. The advantage of the dipolariton system is the freedom in varying the mode energies via an applied electric field, although similar setups could be imagined in triple microcavities~\cite{Diederichs2006}.

Each dispersion mode exhibits a linear polarization splitting between the transverse-electric (TE) and transverse-magnetic (TM) polarizations~\cite{Panzarini1999}. Let us take the TM polarized modes of the middle branch as the relevant quantum modes of our system. A degenerate set of these modes can be distinguished by different orientations of in-plane wavevector and are illustrated in green in Fig.~\ref{fig:ScalableScheme}. In the following we will choose to work in the frame rotating at the energy of this set of modes.

The Hamiltonian of the multimode system up to an energy shift reads (see details in \cite{SM}, sec. 3):
\begin{align}
&\hat{\mathcal{H}}= A \psi_{\mathrm{UP},+} \psi_{\mathrm{LP},+} \hat{a}_{2,+}^\dagger \hat{a}_{1,+}^\dagger  + B \psi_{\mathrm{LP},-}^{'*} \psi_{\mathrm{LP},-} \hat{a}_{2,-}^\dagger \hat{a}_{1,-} \\ \notag
& + C \psi_{\mathrm{UP},+}\psi_{\mathrm{LP},-} \hat{a}_{2,+}^\dagger \hat{a}_{1,-}^\dagger + D \psi_{\mathrm{LP},-}^{'*}\psi_{\mathrm{LP},-} \hat{a}_{2,+}^\dagger \hat{a}_{1,+} +\mathrm{h.c.}
\end{align}
where $A$, $B$, $C$, and $D$ are effective interaction constants which depend on Hopfield coefficients and bare interactions.

The quantum modes can be re-written in terms of their TE and TM polarized components using the transformation:
\begin{equation}
\left(\begin{array}{c}\hat{a}_{n,+}\hat{a}_{n,-}\end{array}\right)=\frac{1}{\sqrt{2}}\left(\begin{array}{cc}e^{2i\phi_n}&ie^{2i\phi_n}\\e^{-2i\phi_n}&-ie^{-2i\phi_n}\end{array}\right)\left(\begin{array}{c}\hat{a}_{n,\mathrm{TM}}\hat{a}_{n,\mathrm{TE}}\end{array}\right),
\end{equation}
where $\phi_n$ denotes the angle of the mode $n$ in reciprocal space.

To arrange for a CNOT gate between an arbitrary pair of modes, $\hat{a}_{1,\mathrm{TM}}$ and $\hat{a}_{2,\mathrm{TM}}$, we consider a quite specific but fully feasible macroscopic excitation of classical fields in the lower and upper polariton branches, as illustrated in red in Fig.~\ref{fig:ScalableScheme}. In the upper branch, we choose a circularly polarized field $\psi_{\mathrm{UP},+}$ at the same wavevector as $\hat{a}_{1,\mathrm{TM}}$. In the lower branch, we choose a cross-circularly polarized field $\psi_{\mathrm{LP},-}$ with the same wavevector as $\hat{a}_{1,\mathrm{TM}}$, together with a linear polarization (characterized by a superposition of both circular components $\psi_{\mathrm{LP},+}$ and $\psi_{\mathrm{LP},-}$) at the same wavevector as $\hat{a}_{2,\mathrm{TM}}$. Since we have chosen classical fields with the same in-plane wavevectors as $\hat{a}_{1,\mathrm{TM}}$ or $\hat{a}_{2,\mathrm{TM}}$, momentum conservation rules allow only a coupling of the chosen quantum modes by the classical fields. In particular, two types of scattering processes appear, where interactions between parallel spins are considered. First, the $\sigma_+$ polarized component of the pumping provides an effective two-mode squeezing interaction, allowing for the scattering processes indicated by the solid arrows in Fig.~\ref{fig:ScalableScheme} (the reverse scattering process also occurs). Second, the $\sigma_-$ polarized component of the pumping results in a process where the quantum modes are exchanged, by transferring their momentum to the classical field. This process is indicated by the dashed arrows in Fig.~\ref{fig:ScalableScheme} (and again the reverse process also occurs).
Finally, the classical field energies can be chosen such that the relevant scattering processes are resonant only with the TM polarized states. Neglecting off-resonant interactions with TE states (see discussion in \cite{SM}, sec. 4) gives the Hamiltonian:
\begin{equation}
\label{eq:H_fin}
\hat{\mathcal{H}}=iJ\hat{a}^\dagger_{1,\mathrm{TM}}\hat{a}^\dagger_{2,\mathrm{TM}}+ iJ\hat{a}^\dagger_{2,\mathrm{TM}}\hat{a}_{1,\mathrm{TM}}+\mathrm{h.c.},
\end{equation}
where $J$ can be made real by correct choice of the amplitudes and phases of the classical fields (\cite{SM}, sec. 3), allowing for CNOT operations acting on eight different continuous variable modes.

{\it Conclusion.---}We presented a scheme for quantum logic gates based on exciton-polaritons in semiconductor microcavities. Unlike previous schemes, we operate with continuous variables that avoid the necessity of operating with a definite number of polaritons, and make use of an effective amplification of nonlinearity in the system based on the coupling of quantum modes with macroscopically occupied classical states. Using these ingredients, a quantum optical treatment of decay processes predicts fidelities in excess of 99\% for existing microcavities. Additionally, we proposed a way for a construction of scalable networks of polaritonic gates. The experimental demonstration of our proposal would not be reliant on single-photon detection and would require standard homodyne detection measurements (this would depend also on the implementation of future error correction protocols).


\clearpage

\setcounter{equation}{0} \setcounter{figure}{0} \renewcommand{%
\theequation}{S\arabic{equation}} \renewcommand{\thefigure}{S\arabic{figure}}

\noindent {\bf 1. Possible implementation of an experimental detection scheme}\\

\noindent The characterization of a quantum optical process can be achieved by measuring its effect on input coherent states~\cite{Lobino2008}. Gaussian coherent states are fully characterized by their covariance matrix, which can be experimentally accessed using homodyne detection~\cite{Auria2009}. An important ingredient of such a technique is the availability of a classical local oscillator, with identical frequency to the measured modes. While this is not immediately available in our system, a modified detection scheme can be implemented using an additional interference between the lasers driving upper and lower branch polaritons.

For the described system, the relevant two-mode covariance matrix is given by
\begin{equation}
V_{ij}=\frac{1}{2}\left<\hat{x}_i\hat{x}_j+\hat{x}_j\hat{x}_i\right>-\left<\hat{x}_i\right>\left<\hat{x}_j\right>,
\end{equation}
where $\hat{x}^T=\left(\hat{q}_1,\hat{p}_1,\hat{q}_2,\hat{p}_2\right)$.

\begin{figure}
\includegraphics[width=\linewidth]{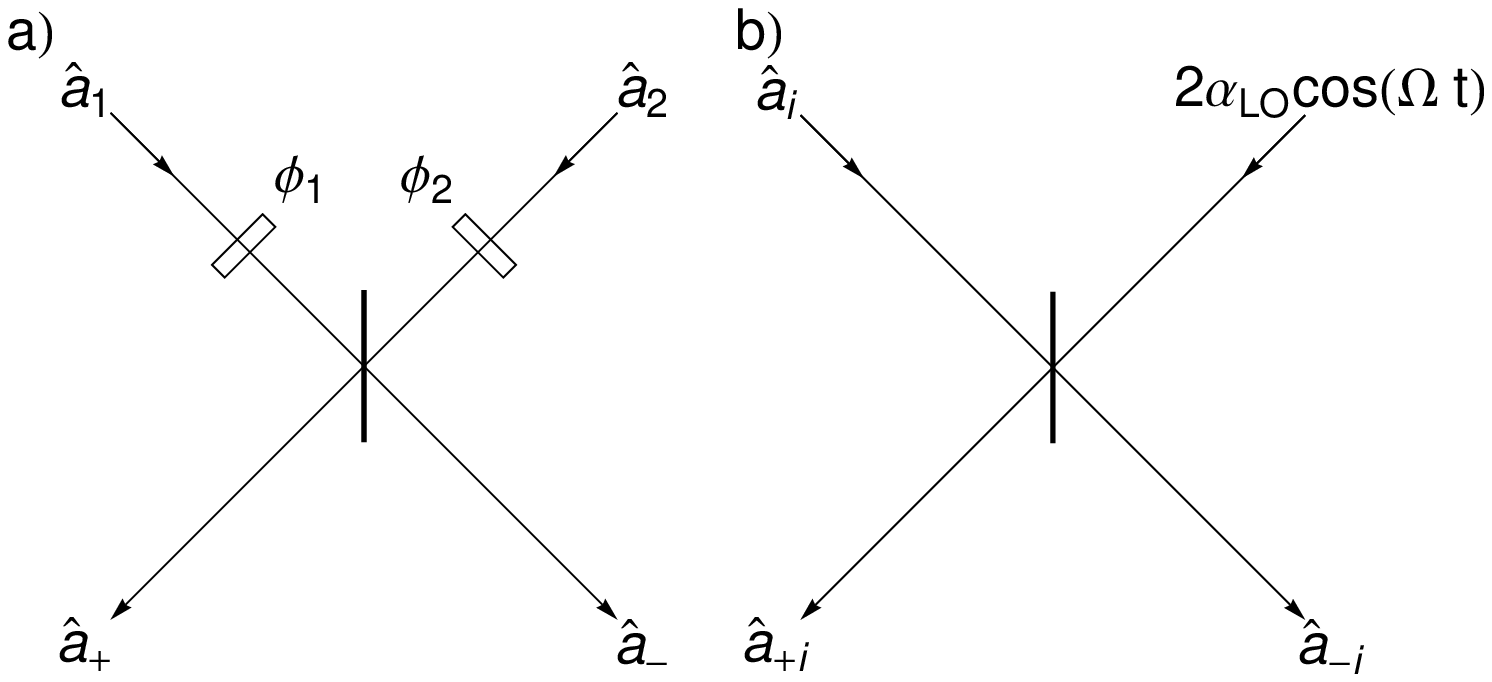}
\caption{Experimental scheme for measurement of covariance matrix. a) Using phase delays ($\phi_1$, $\phi_2$) and interfering on a beam-splitter generates the fields $\hat{a}_\pm=\left(\hat{a}_1e^{i\phi_1}\pm\hat{a}_2e^{i\phi_2}\right)/\sqrt{2}$. Different combinations of phases then allow access to the fields $\hat{a}_i$ as defined in the text. b) Interfering $\hat{a}_i$ with a local oscillator of the form $\alpha_\mathrm{LO}\left(e^{i\Omega t}+e^{-i\Omega t}\right)$ allows generation of the fields $\hat{a}_{\pm i}=\left[\hat{a}_i\pm\alpha_\mathrm{LO}\left(e^{i\Omega t}+e^{-i\Omega t}\right)\right]/\sqrt{2}$. The differences of the corresponding photocurrents gives access to the quadratures of $\hat{a}_i$.}
\label{fig:HeterodyneScheme}
\end{figure}

From the interference of the modes $\hat{a}_1$ and $\hat{a}_2$ on a beam-splitter and the application of phase delays, one has access to the fields $\hat{a}_3=(\hat{a}_1+\hat{a}_2)/\sqrt{2}$, $\hat{a}_4=(\hat{a}_1-\hat{a}_2)/\sqrt{2}$, $\hat{a}_5=(i\hat{a}_1+\hat{a}_2)/\sqrt{2}$, $\hat{a}_6=(i\hat{a}_1-\hat{a}_2)/\sqrt{2}$ [see Fig.~\ref{fig:HeterodyneScheme}(a)].

The fields $\hat{a}_i$ all oscillate at the same frequency, which is midway between the frequency of the laser driving the upper polariton branch and the frequency of the laser driving the lower polariton branch. Let us consider interfering any of the fields $\hat{a}_i$ on a beam-splitter with a local oscillator field of the form $\alpha_\mathrm{LO}\left(e^{i\Omega t}+e^{-i\Omega t}\right)$. This form of local oscillator can be attained from the superposition of the lasers driving the lower and upper branches, where $\alpha_\mathrm{LO}$ is the (complex) field amplitude and $\Omega$ is the frequency difference with the middle polariton modes. The photocurrents obtained at each of the output ports are [see Fig.~\ref{fig:HeterodyneScheme}(b)]:
\begin{align}
\hat{n}_{+i}&=\frac{1}{2}\left(\hat{a}^\dagger_i+\alpha_\mathrm{LO}^*\left(e^{i\Omega t}+e^{-i\Omega t}\right)\right)\notag\\
&\hspace{20mm}\times\left(\hat{a}_i+\alpha_\mathrm{LO}\left(e^{i\Omega t}+e^{-i\Omega t}\right)\right),\\
\hat{n}_{-i}&=\frac{1}{2}\left(\hat{a}^\dagger_i-\alpha_\mathrm{LO}^*\left(e^{i\Omega t}+e^{-i\Omega t}\right)\right)\notag\\
&\hspace{20mm}\times\left(\hat{a}_i-\alpha_\mathrm{LO}\left(e^{i\Omega t}+e^{-i\Omega t}\right)\right).
\end{align}

The difference photocurrent is then:
\begin{equation}
\hat{n}_{+i}-\hat{n}_{-i}=2\left(\hat{a}^\dagger_i\alpha_\mathrm{LO}+\hat{a}_i\alpha_\mathrm{LO}^*\right)\cos\left(\Omega t\right).
\end{equation}

The $\cos\left(\Omega t\right)$ modulation can be identified and demodulated. By varying the phase of $\alpha_\mathrm{LO}$ one can access $\langle\hat{a}^\dagger_i\pm\hat{a}_i\rangle$ from the average values and $\langle\left(\hat{a}^\dagger_i\pm\hat{a}_i\right)^2\rangle$ from the observed variation. From these quantities, the covariance matrix can be reconstructed as in Ref.~\cite{Auria2009}.\\

\noindent {\bf 2. Analytic solution of the first and second order correlator evolution}\\

\noindent To calculate the fidelity of polaritonic CNOT gate as a function of decay rate $\Gamma$ and pure dephasing rate $\Gamma_{P}$, we derive the analytic solutions for amplitude and phase expectation values, as wells as higher order correlators. The solution of Eqs.~(4)-(7) in the main text reads:
\begin{align}
\langle\hat{q}_1(t)\rangle&=\langle\hat{q}_1(0)\rangle e^{(P-\Gamma-\Gamma_{P})t/(2\hbar)},\\
\langle\hat{q}_2(t)\rangle&=e^{(P-\Gamma-\Gamma_{P})t/(2\hbar)}\left(\langle\hat{q}_2(0)\rangle+\frac{2J}{\hbar}\langle\hat{q}_1(0)\rangle t\right),\\
\langle\hat{p}_1(t)\rangle&=e^{(P-\Gamma-\Gamma_{P})t/(2\hbar)}\left(\langle\hat{p}_1(0)\rangle-\frac{2J}{\hbar}\langle\hat{p}_2(0)\rangle t\right),\\
\langle \hat{p}_2(t)\rangle&=\langle\hat{p}_2(0)\rangle e^{(P-\Gamma-\Gamma_{P}) t/(2\hbar)},
\end{align}
where $\langle\hat{q}_n(0)\rangle$ and $\langle\hat{p}_n(0)\rangle$ represent the initial amplitude and phase mean-field values.

The evolution of second order correlators can be derived from Eq.~(3) as:
\begin{align}
\label{eq:dq1q1dt}
i\hbar\frac{d\langle \hat{q}_1^2\rangle}{dt}&=\frac{i\Gamma}{2}\left(1-2\langle\hat{q}_1^2\rangle\right)+\frac{iP}{2}\left(1+2\langle\hat{q}_1^2\rangle\right) \\ \notag &\hspace{3mm}-i\Gamma_{P}\left(\langle\hat{q}_1^2\rangle - \langle\hat{p}_1^2\rangle\right),\\
i\hbar\frac{d\langle \hat{q}_1\hat{q}_2\rangle}{dt}&=2iJ\langle\hat{q}_1^2\rangle+i(P-\Gamma-2\Gamma_{P})\langle\hat{q}_1\hat{q}_2\rangle,\\
\label{eq:dq2q2dt}
i\hbar\frac{d\langle \hat{q}_2^2\rangle}{dt}&=4iJ\langle\hat{q}_1\hat{q}_2\rangle+\frac{i\Gamma}{2}\left(1-2\langle\hat{q}_2^2\rangle\right)\\ \notag
&\hspace{3mm}+\frac{iP}{2}\left(1+2\langle\hat{q}_2^2\rangle\right)-i\Gamma_{P}\left(\langle\hat{q}_2^2\rangle - \langle\hat{p}_2^2\rangle\right),
\end{align}
\begin{align}
\label{eq:dp1p1dt}
i\hbar\frac{d\langle \hat{p}_1^2\rangle}{dt}&=-4iJ\langle\hat{p}_1\hat{p}_2\rangle+\frac{i\Gamma}{2}\left(1-2\langle\hat{p}_1^2\rangle\right)\\ \notag
&\hspace{3mm}+\frac{iP}{2}\left(1+2\langle\hat{p}_1^2\rangle\right)-i\Gamma_{P}\left(\langle\hat{p}_1^2\rangle - \langle\hat{q}_1^2\rangle\right),\\
i\hbar\frac{d\langle \hat{p}_1\hat{p}_2\rangle}{dt}&=-2iJ\langle\hat{p}_2^2\rangle+i(P-\Gamma-2\Gamma_{P})\langle\hat{p}_1\hat{p}_2\rangle,\\
\label{eq:dp2p2dt}
i\hbar\frac{d\langle \hat{p}_2^2\rangle}{dt}&=\frac{i\Gamma}{2}\left(1-2\langle\hat{p}_2^2\rangle\right)+\frac{iP}{2}\left(1+2\langle\hat{p}_2^2\rangle\right) \\ \notag &\hspace{3mm}-i\Gamma_{P}\left(\langle\hat{p}_2^2\rangle - \langle\hat{q}_2^2\rangle\right),
\end{align}
\begin{align}
i\hbar\frac{\langle\hat{q}_1\hat{p}_2\rangle}{dt}&=i\left(P-\Gamma-2\Gamma_{P}\right)\langle\hat{q}_1\hat{p}_2\rangle,\\
i\hbar\frac{\langle\hat{q}_1\hat{p}_1\rangle}{dt}&=-2iJ\langle\hat{q}_1\hat{p}_2\rangle+\frac{i(P-\Gamma-2\Gamma_{P})}{2}\left(2\langle\hat{q}_1\hat{p}_1\rangle-i\right),\\
i\hbar\frac{\langle\hat{q}_2\hat{p}_1\rangle}{dt}&=2iJ\left(\langle\hat{q}_1\hat{p}_1\rangle-\langle\hat{q}_2\hat{p}_2\rangle\right)+i\left(P-\Gamma-2\Gamma_{P}\right)\langle\hat{q}_2\hat{p}_1\rangle,\\
i\hbar\frac{\langle\hat{q}_2\hat{p}_2\rangle}{dt}&=2iJ\langle\hat{q}_1\hat{p}_2\rangle+\frac{i(P-\Gamma-2\Gamma_{P})}{2}\left(2\langle\hat{q}_2\hat{p}_2\rangle-i\right).\label{eq:dq2p2dt}
\end{align}

Eq.~(\ref{eq:dq1q1dt})-(\ref{eq:dq2p2dt}) can be solved analytically for $\Gamma_{P} = 0$ and $\Gamma\neq P$, giving:

\begin{align}
\langle\hat{q}^2_1(t)\rangle&=c_1+\left(\langle\hat{q}^2_1(0)\rangle-c_1\right)e^{(P-\Gamma)t/\hbar},\\
\langle\hat{q}_1(t)\hat{q}_2(t)\rangle&=c_2+\left(\langle\hat{q}_1(0)\hat{q}_2(0)\rangle-c_2\right)e^{(P-\Gamma)t/\hbar}\notag\\
&\hspace{5mm}+\left(c_3+\frac{2J}{\hbar}\langle\hat{q}^2_1(0)\rangle\right)te^{(P-\Gamma)t/\hbar},\\
\langle\hat{q}^2_2(t)\rangle&=c_4+\left(\langle\hat{q}^2_2(0)\rangle-c_4+\frac{4J\left(\langle\hat{q}_1\hat{q}_2(0)\rangle-c_2\right)}{\hbar}t\right.\notag\\
&\hspace{5mm}\left.+\frac{2J\left(c_3\hbar+2J\langle\hat{q}^2_1(0)\rangle\right)}{\hbar^2}t^2\right)e^{(P-\Gamma)t/\hbar},
\end{align}
\begin{align}
\langle\hat{p}^2_2(t)\rangle&=c_1+\left(\langle\hat{p}^2_2(0)\rangle-c_1\right)e^{(P-\Gamma)t/\hbar},\\
\langle\hat{p}_1(t)\hat{p}_2(t)\rangle&=-c_2+\left(\langle\hat{p}_1(0)\hat{p}_2(0)\rangle+c_2\right)e^{(P-\Gamma)t/\hbar}\notag\\
&\hspace{5mm}-\left(c_3+\frac{2J}{\hbar}\langle\hat{p}^2_2(0)\rangle\right)te^{(P-\Gamma)t/\hbar},\\
\langle\hat{p}^2_1(t)\rangle&=c_4+\left(\langle\hat{p}^2_1(0)\rangle-c_4-\frac{4J\left(\langle\hat{p}_1\hat{p}_2(0)\rangle+c_2\right)}{\hbar}t\right.\notag\\
&\hspace{5mm}\left.+\frac{2J\left(c_3\hbar+2J\langle\hat{p}^2_2(0)\rangle\right)}{\hbar^2}t^2\right)e^{(P-\Gamma)t/\hbar},
\end{align}
\begin{align}
\langle\hat{q}_1\hat{p}_2(t)\rangle&=\langle\hat{q}_1\hat{p}_2(0)\rangle e^{(P-\Gamma)t/\hbar},\\
\langle\hat{q}_1\hat{p}_1(t)\rangle&=\notag\\
\frac{i}{2}&+\left(\langle\hat{q}_1\hat{p}_1(0)-\frac{i}{2}-\frac{2J\langle\hat{q}_1\hat{p}_2(0)\rangle t}{\hbar}\right)e^{(P-\Gamma)t/\hbar},\\
\langle\hat{q}_2\hat{p}_2(t)\rangle&=\notag\\
\frac{i}{2}&+\left(\langle\hat{q}_2\hat{p}_2(0)-\frac{i}{2}+\frac{2J\langle\hat{q}_1\hat{p}_2(0)\rangle t}{\hbar}\right)e^{(P-\Gamma)t/\hbar},\\
\langle\hat{q}_2\hat{p}_1(t)\rangle&=\left(\langle\hat{q}_2\hat{p}_1(0)\rangle+\frac{2J}{\hbar}\left(\langle\hat{q}_1\hat{p}_1(0)\rangle-\langle\hat{q}_2\hat{p}_2(0)\rangle\right)t\right.\notag\\
&\hspace{15mm}\left.-\frac{4J^2\langle\hat{q}_1\hat{p}_2(0)\rangle t^2}{\hbar^2}\right)e^{(P-\Gamma)t/\hbar},
\end{align}
where the coefficients are:
\begin{align}
c_1&=\frac{P+\Gamma}{2(\Gamma-P)},\\
c_2&=\frac{2c_1J}{\Gamma-P},\\
c_3&=-2c_1\frac{J}{\hbar},\\
c_4&=c_1+\frac{4c_2J}{\Gamma-P}.
\end{align}
Separate equations hold for the special case $\Gamma=P$.\\



\noindent {\bf 3. Multi gate realization: detailed description }\\

In the main text of the letter we described the scheme for a multi-CNOT gate, which can operate on different pairs of continuous variable modes, encoded by the planar wavevector. The following can be done for instance in the spinfor dipolaritonic system, where direct exciton (DX), indirect exciton (IX) and cavity photon (C) form three distinct dipolaritonic modes UP, MP, and LP, with associated parametric scattering between them. Alternatively, one can envisage a similar system, where two different direct exciton modes DX$_1$ and DX$_2$ are coupled to the same cavity mode C.

We start the description from a generic dipolaritonic Hamiltonian \cite{Kyriienko2014SM,Byrnes2014}
\begin{equation}
\label{eq:H_dip}
\hat{\mathcal{H}}_{\mathrm{dip}} = \hat{\mathcal{H}}_{0} + \hat{\mathcal{H}}_{\mathrm{int}},
\end{equation}
where we separated the linear Hamiltonian of bare modes and associated couplings, $\hat{\mathcal{H}}_{0}$, and interaction Hamiltonian $\hat{\mathcal{H}}_{\mathrm{int}}$ coming from exciton-exciton interactions.

The linear part reads
\begin{align}
\label{eq:H_0}
&\hat{\mathcal{H}}_{0} = \sum\limits_{\mathbf{k},j=\pm} E_{\mathrm{C},\mathbf{k},j} \hat{a}_{\mathbf{k},j}^\dagger \hat{a}_{\mathbf{k},j} + \sum\limits_{\mathbf{k},j=\pm} E_{\mathrm{DX},\mathbf{k},j} \hat{b}_{\mathbf{k},j}^\dagger \hat{b}_{\mathbf{k},j} \\
\notag &+ \sum\limits_{\mathbf{k},j=\pm} E_{\mathrm{IX},\mathbf{k},j} \hat{c}_{\mathbf{k},j}^\dagger \hat{c}_{\mathbf{k},j} + \sum\limits_{\mathbf{k},j=\pm} \Omega_{\mathrm{C-DX}} (\hat{a}_{\mathbf{k},j}^\dagger \hat{b}_{\mathbf{k},j} + \hat{b}_{\mathbf{k},j}^\dagger \hat{a}_{\mathbf{k},j}) \\
\notag &- \sum\limits_{\mathbf{k},j=\pm} J_{\mathrm{DX-IX}} (\hat{a}_{\mathbf{k},j}^\dagger \hat{b}_{\mathbf{k},j} + \hat{b}_{\mathbf{k},j}^\dagger \hat{a}_{\mathbf{k},j}),
\end{align}
where $\hat{a}_{\mathbf{k},j}$ ($\hat{a}_{\mathbf{k},j}^\dagger$), $\hat{b}_{\mathbf{k},j}$ ($\hat{b}_{\mathbf{k},j}^\dagger$), $\hat{c}_{\mathbf{k},j}$ ($\hat{c}_{\mathbf{k},j}^\dagger$) correspond to annihilation (creation) operators of the cavity photon, direct exciton, and indirect exciton modes with $\mathbf{k}$ planar wave vector and $j = \pm$ circular polarization, respectively. The dispersions of the C, DX, and IX modes are $E_{\mathrm{C},\mathbf{k},j}$, $E_{\mathrm{DX},\mathbf{k},j}$, and $E_{\mathrm{IX},\mathbf{k},j}$, respectively, and we consider circular modes to be degenerate in energy, $E_{\cdot,\mathbf{k},+} = E_{\cdot,\mathbf{k},-}$. $\Omega_{\mathrm{C-DX}}$ corresponds to the exciton-photon coupling constant, and $J_{\mathrm{DX-IX}}$ denotes tunneling coupling between adjacent quantum wells.

The interaction Hamiltonian can be written as
\begin{equation}
\label{eq:H_int}
\hat{\mathcal{H}}_{\mathrm{int}} = \hat{\mathcal{H}}_{\mathrm{DX-DX}} + \hat{\mathcal{H}}_{\mathrm{IX-IX}} + \hat{\mathcal{H}}_{\mathrm{DX-IX}},
\end{equation}
where we separate contributions emerging from direct exciton interaction, indirect exciton interaction, and direct-indirect cross-Kerr interaction. Starting the from conventional DX-DX interaction, the Hamiltonian reads
\begin{align}
\notag
&\hat{\mathcal{H}}_{\mathrm{DX-DX}} = \sum\limits_{\mathbf{k},\mathbf{k'},\mathbf{q}} \alpha_{1}^{\mathrm{DD}}(\mathbf{k},\mathbf{k'},\mathbf{q}) \hat{b}_{\mathbf{k-q},+}^\dagger \hat{b}_{\mathbf{k'+q},+}^\dagger \hat{b}_{\mathbf{k},+} \hat{b}_{\mathbf{k'},+} \\
\label{eq:H_DX-DX}
&+ \sum\limits_{\mathbf{k},\mathbf{k'},\mathbf{q}} \alpha_{1}^{\mathrm{DD}}(\mathbf{k},\mathbf{k'},\mathbf{q}) \hat{b}_{\mathbf{k-q},-}^\dagger \hat{b}_{\mathbf{k'+q},-}^\dagger \hat{b}_{\mathbf{k},-} \hat{b}_{\mathbf{k'},-} \\ \notag
&+ \sum\limits_{\mathbf{k},\mathbf{k'},\mathbf{q}} \alpha_{2}^{\mathrm{DD}}(\mathbf{k},\mathbf{k'},\mathbf{q}) \hat{b}_{\mathbf{k-q},+}^\dagger \hat{b}_{\mathbf{k},+} \hat{b}_{\mathbf{k'+q},-}^\dagger \hat{b}_{\mathbf{k'},-} \\ \notag
&+ \sum\limits_{\mathbf{k},\mathbf{k'},\mathbf{q}} \alpha_{2}^{\mathrm{DD}}(\mathbf{k},\mathbf{k'},\mathbf{q}) \hat{b}_{\mathbf{k-q},-}^\dagger \hat{b}_{\mathbf{k},-} \hat{b}_{\mathbf{k'+q},+}^\dagger \hat{b}_{\mathbf{k'},+},
\end{align}
where $\alpha_{1}^{\mathrm{DD}}$ corresponds to the triplet or co-circular Coulomb interaction between direct excitons of the same spin, and $\alpha_{2}^{\mathrm{DD}}$ is the singlet or cross-circular interaction between direct excitons of opposite polarization. We note that for relevant momenta of the polaritonic system the exciton-exciton interaction constant is momentum independent and can be typically treated as a constant.

In the following we are interested in particular wave vectors of the particles, namely the one where scattering happens between initial states of $\mathbf{k}_2$ and $\mathbf{k}_1$, and final states $\mathbf{k'}_1$ and $\mathbf{k'}_2$ chosen according to momentum conservation. Using this labelling, Eq. (\ref{eq:H_DX-DX}) yields
\begin{align}
\notag
\hat{\mathcal{H}}_{\mathrm{DX-DX}} &= \alpha_{1}^{\mathrm{DD}} \hat{b}_{\mathbf{k}_2',+}^\dagger \hat{b}_{\mathbf{k}_1',+}^\dagger \hat{b}_{\mathbf{k}_2,+} \hat{b}_{\mathbf{k}_1,+} \\
\label{eq:H_DX-DX_2}
&+ \alpha_{1}^{\mathrm{DD}} \hat{b}_{\mathbf{k}_2',-}^\dagger \hat{b}_{\mathbf{k}_1',-}^\dagger \hat{b}_{\mathbf{k}_2,-} \hat{b}_{\mathbf{k}_1,-} \\ \notag
&+ \alpha_{2}^{\mathrm{DD}} \hat{b}_{\mathbf{k}_2',+}^\dagger \hat{b}_{\mathbf{k}_2,+} \hat{b}_{\mathbf{k}_1',-}^\dagger \hat{b}_{\mathbf{k}_1,-} \\ \notag
&+ \alpha_{2}^{\mathrm{DD}} \hat{b}_{\mathbf{k}_2',-}^\dagger \hat{b}_{\mathbf{k}_2,-} \hat{b}_{\mathbf{k}_1',+}^\dagger \hat{b}_{\mathbf{k}_1,+}.
\end{align}
The interaction terms for indirect excitons can be written in a similar fashion, giving
\begin{align}
\notag
\hat{\mathcal{H}}_{\mathrm{IX-IX}} &= \alpha_{1}^{\mathrm{II}} \hat{c}_{\mathbf{k}_2',+}^\dagger \hat{c}_{\mathbf{k}_1',+}^\dagger \hat{c}_{\mathbf{k}_2,+} \hat{c}_{\mathbf{k}_1,+} \\
\label{eq:H_IX-IX_2}
&+ \alpha_{1}^{\mathrm{II}} \hat{c}_{\mathbf{k}_2',-}^\dagger \hat{c}_{\mathbf{k}_1',-}^\dagger \hat{c}_{\mathbf{k}_2,-} \hat{c}_{\mathbf{k}_1,-} \\ \notag
&+ \alpha_{2}^{\mathrm{II}} \hat{c}_{\mathbf{k}_2',+}^\dagger \hat{c}_{\mathbf{k}_2,+} \hat{c}_{\mathbf{k}_1',-}^\dagger \hat{c}_{\mathbf{k}_1,-} \\ \notag
&+ \alpha_{2}^{\mathrm{II}} \hat{c}_{\mathbf{k}_2',-}^\dagger \hat{c}_{\mathbf{k}_2,-} \hat{c}_{\mathbf{k}_1',+}^\dagger \hat{c}_{\mathbf{k}_1,+},
\end{align}
with $\alpha_{1,2}^{\mathrm{II}}$ triplet/singlet interaction between spinor indirect excitons.
The cross-interaction between direct and indirect excitons reads
\begin{align}
\notag
\hat{\mathcal{H}}_{\mathrm{DX-IX}} &= \alpha_{1}^{\mathrm{DI}} \hat{b}_{\mathbf{k}_2',+}^\dagger \hat{c}_{\mathbf{k}_1',+}^\dagger \hat{b}_{\mathbf{k}_2,+} \hat{c}_{\mathbf{k}_1,+} \\
\label{eq:H_DX-IX_2}
&+ \alpha_{1}^{\mathrm{DI}} \hat{b}_{\mathbf{k}_2',-}^\dagger \hat{c}_{\mathbf{k}_1',-}^\dagger \hat{b}_{\mathbf{k}_2,-} \hat{c}_{\mathbf{k}_1,-} \\ \notag
&+ \alpha_{2}^{\mathrm{DI}} \hat{b}_{\mathbf{k}_2',+}^\dagger \hat{b}_{\mathbf{k}_2,+} \hat{c}_{\mathbf{k}_1',-}^\dagger \hat{c}_{\mathbf{k}_1,-} \\ \notag
&+ \alpha_{2}^{\mathrm{DI}} \hat{b}_{\mathbf{k}_2',-}^\dagger \hat{b}_{\mathbf{k}_2,-} \hat{c}_{\mathbf{k}_1',+}^\dagger \hat{c}_{\mathbf{k}_1,+},
\end{align}
where $\alpha_{1,2}^{\mathrm{DI}}$ corresponds to direct-indirect exciton Coulomb interaction for same and opposite spin projections $\pm 1$.

Next, we perform the transformation of the linear Hamiltonian (\ref{eq:H_0}) to the diagonal basis of dipolariton operators $\hat{A}_j$ ($j=1,2,3$):
\begin{align}
\hat{a}_{k,\pm}&=V_{11,k,\pm}\hat{A}_{1,k,\pm}+V_{21,k,\pm}\hat{A}_{2,k,\pm}+V_{31,k,\pm}\hat{A}_{3,k,\pm}, \label{eq:a}\\
\hat{b}_{k,\pm}&=V_{12,k,\pm}\hat{A}_{1,k,\pm}+V_{22,k,\pm}\hat{A}_{2,k,\pm}+V_{32,k,\pm}\hat{A}_{3,k,\pm}, \label{eq:b}\\
\hat{c}_{k,\pm}&=V_{13,k,\pm}\hat{A}_{1,k,\pm}+V_{23,k,\pm}\hat{A}_{2,k,\pm}+V_{33,k,\pm}\hat{A}_{3,k,\pm}, \label{eq:c}
\end{align}
where $V_{ij}$ are the matrix elements of eigenvectors, corresponding to dipolariton Hopfield coefficients. Without limiting the generality we consider $V_{ij}$ to be real for any $i$ and $j$.

The transformed linear part reads
\begin{equation}
\hat{\mathcal{H}}_0^{'} = \sum\limits_{j, \mathbf{k}, \pm} E_{j,\mathbf{k},\pm} \hat{A}_{j,\mathbf{k},\pm}^\dagger \hat{A}_{j,\mathbf{k},\pm}.
\end{equation}
The eigenenergies possess rotation symmetry, $E_j(\mathbf{k}) = E_j(k)$, and thus the Hopfield coefficients are spin independent, $V_{ij,\mathbf{k},\pm} \equiv V_{ij}$, where we omit the momentum index for brevity.

We proceed with the transformation of the interaction Hamiltonian by successive transformation of its parts. For instance, the direct exciton interaction part can be rewritten as
\begin{widetext}
\begin{align}
\label{eq:H_DX-DX_prime}
&\hat{\mathcal{H}}_{\mathrm{DX-DX}}^{'} = \alpha_{1}^{\mathrm{DD}} (V_{12}\hat{A}_{1,\mathbf{k}_2',+}^\dagger + V_{22}\hat{A}_{2,\mathbf{k}_2',+}^\dagger + V_{32}\hat{A}_{3,\mathbf{k}_2',+}^\dagger) (V_{12}\hat{A}_{1,\mathbf{k}_1',+}^\dagger + V_{22}\hat{A}_{2,\mathbf{k}_1',+}^\dagger + V_{32}\hat{A}_{3,\mathbf{k}_1',+}^\dagger) \\ \notag &(V_{12}\hat{A}_{1,\mathbf{k}_2,+} + V_{22}\hat{A}_{2,\mathbf{k}_2,+} + V_{32}\hat{A}_{3,\mathbf{k}_2,+}) (V_{12}\hat{A}_{1,\mathbf{k}_1,+} + V_{22}\hat{A}_{2,\mathbf{k}_1,+} + V_{32}\hat{A}_{3,\mathbf{k}_1,+}) +  \alpha_{1}^{\mathrm{DD}} (V_{12}\hat{A}_{1,\mathbf{k}_2',-}^\dagger \\ \notag &+ V_{22}\hat{A}_{2,\mathbf{k}_2',-}^\dagger + V_{32}\hat{A}_{3,\mathbf{k}_2',-}^\dagger) (V_{12}\hat{A}_{1,\mathbf{k}_1',-}^\dagger + V_{22}\hat{A}_{2,\mathbf{k}_1',-}^\dagger + V_{32}\hat{A}_{3,\mathbf{k}_1',-}^\dagger) (V_{12}\hat{A}_{1,\mathbf{k}_2,-} + V_{22}\hat{A}_{2,\mathbf{k}_2,-} + V_{32}\hat{A}_{3,\mathbf{k}_2,-}) \\ \notag &(V_{12}\hat{A}_{1,\mathbf{k}_1,-} + V_{22}\hat{A}_{2,\mathbf{k}_1,-} + V_{32}\hat{A}_{3,\mathbf{k}_1,-}) + \alpha_{2}^{\mathrm{DD}} (V_{12}\hat{A}_{1,\mathbf{k}_2',+}^\dagger + V_{22}\hat{A}_{2,\mathbf{k}_2',+}^\dagger + V_{32}\hat{A}_{3,\mathbf{k}_2',+}^\dagger) \\ \notag & (V_{12}\hat{A}_{1,\mathbf{k}_2,+} + V_{22}\hat{A}_{2,\mathbf{k}_2,+} + V_{32}\hat{A}_{3,\mathbf{k}_2,+}) (V_{12}\hat{A}_{1,\mathbf{k}_1',-}^\dagger + V_{22}\hat{A}_{2,\mathbf{k}_1',-}^\dagger + V_{32}\hat{A}_{3,\mathbf{k}_1',-}^\dagger) (V_{12}\hat{A}_{1,\mathbf{k}_1,-} + V_{22}\hat{A}_{2,\mathbf{k}_1,-} + V_{32}\hat{A}_{3,\mathbf{k}_1,-}) \\ \notag
& + \alpha_{2}^{\mathrm{DD}} (V_{12}\hat{A}_{1,\mathbf{k}_2',-}^\dagger + V_{22}\hat{A}_{2,\mathbf{k}_2',-}^\dagger + V_{32}\hat{A}_{3,\mathbf{k}_2',-}^\dagger) (V_{12}\hat{A}_{1,\mathbf{k}_2,-} + V_{22}\hat{A}_{2,\mathbf{k}_2,-} + V_{32}\hat{A}_{3,\mathbf{k}_2,-}) \\ \notag &(V_{12}\hat{A}_{1,\mathbf{k}_1',+}^\dagger + V_{22}\hat{A}_{2,\mathbf{k}_1',+}^\dagger + V_{32}\hat{A}_{3,\mathbf{k}_1',+}^\dagger) (V_{12}\hat{A}_{1,\mathbf{k}_1,+} + V_{22}\hat{A}_{2,\mathbf{k}_1,+} + V_{32}\hat{A}_{3,\mathbf{k}_1,+}).
\end{align}
\end{widetext}
To proceed, we recall the relevant quantum and classical modes required for the operation of a serial multimode CNOT gate shown for example in Fig. 3, main text. Associating the UP, MP, and LP modes with $\hat{A}_1$, $\hat{A}_2$, and $\hat{A}_3$ modes, the generic interaction Hamiltonian (\ref{eq:H_DX-DX_prime}) can be accommodated to our needs using the set of relabellings:
\begin{align}
\label{eq:labels}
&\hat{A}_{1,\mathbf{k}_1,+}\mapsto\psi_{\mathrm{UP},+}, \quad \hat{A}_{1,\mathbf{k}_1,-}\mapsto 0, \\ \notag &\hat{A}_{1,\mathbf{k}_2,+}\mapsto 0, \quad \hat{A}_{1,\mathbf{k}_2,-}\mapsto 0,\\ \notag
&\hat{A}_{2,\mathbf{k}_1,+}\mapsto \hat{a}_{1,+}, \quad \hat{A}_{2,\mathbf{k}_1,-}\mapsto \hat{a}_{1,-}, \\ \notag &\hat{A}_{2,\mathbf{k}_2,+}\mapsto \hat{a}_{2,+}, \quad \hat{A}_{2,\mathbf{k}_2,-}\mapsto \hat{a}_{2,-}, \\ \notag &\hat{A}_{3,\mathbf{k}_1,+}\mapsto 0,\quad \hat{A}_{3,\mathbf{k}_1,-}\mapsto \psi_{\mathrm{LP},-}^{'}, \\ \notag &\hat{A}_{3,\mathbf{k}_2,+}\mapsto \psi_{\mathrm{LP},+}, \quad \hat{A}_{3,\mathbf{k}_2,-}\mapsto \psi_{\mathrm{LP},-},
\end{align}
and again the final state (primed) wavevectors are chosen according to momentum conservation rules. These imply that $\mathbf{k}_1 + \mathbf{k}_2 = \mathbf{k}_2' + \mathbf{k}_1'$ for parametric coupling processes and $\mathbf{k}_2 - \mathbf{k}_1 = \mathbf{k}_2' - \mathbf{k}_1'$ for linear coupling processes, which can always be satisfied by choosing wave vectors for the classical modes appropriately.

Using Eq. (\ref{eq:labels}) we can rewrite Eq. (\ref{eq:H_DX-DX_prime}) as
\begin{widetext}
\begin{align}
\label{eq:H_DX-DX_prime_mod}
&\hat{\mathcal{H}}_{\mathrm{DX-DX}}^{'} = \alpha_{1}^{\mathrm{DD}} (V_{22}\hat{a}_{2,+}^\dagger + V_{32}\psi_{\mathrm{LP},+}^{*}) (V_{12} \psi_{\mathrm{UP},+}^{*} + V_{22}\hat{a}_{1,+}^\dagger) (V_{12}\psi_{\mathrm{UP},+} + V_{22}\hat{a}_{1,+}) (V_{22}\hat{a}_{2,+} + V_{32}\psi_{\mathrm{LP},+}) \\ \notag &+ \alpha_{1}^{\mathrm{DD}} (V_{22}\hat{a}_{2,-}^\dagger + V_{32}\psi_{\mathrm{LP},-}^{*}) (V_{22}\hat{a}_{1,-}^\dagger + V_{32}\psi_{\mathrm{LP},-}^{'*}) (V_{22}\hat{a}_{1,-} + V_{32}\psi_{\mathrm{LP},-}^{'}) (V_{22}\hat{a}_{2,-} + V_{32}\psi_{\mathrm{LP},-}) \\ \notag
& + \alpha_{2}^{\mathrm{DD}} (V_{22}\hat{a}_{2,+}^\dagger + V_{32}\psi_{\mathrm{LP},+}^{*})  (V_{12}\psi_{\mathrm{UP},+} + V_{22}\hat{a}_{1,+}) (V_{22}\hat{A}_{1,-}^\dagger + V_{32}\psi_{\mathrm{LP},-}^{'*}) (V_{22}\hat{A}_{2,-} + V_{32}\psi_{\mathrm{LP},-}) \\ \notag
& + \alpha_{2}^{\mathrm{DD}} (V_{22}\hat{a}_{2,-}^\dagger + V_{32}\psi_{\mathrm{LP},-}^{*}) (V_{22}\hat{a}_{1,-} + V_{32}\psi_{\mathrm{LP},-}^{'}) (V_{12}\psi_{\mathrm{UP},+}^{*} + V_{22}\hat{a}_{1,+}^\dagger) (V_{22}\hat{a}_{2,+} + V_{32}\psi_{\mathrm{LP},+}).
\end{align}
\end{widetext}
Taking only energy conserving terms, we arrive to the Hamiltonian derived from the DX-DX interaction:
\begin{align}
\label{eq:H_DX-DX_RWA}
&\hat{\mathcal{H}}_{\mathrm{DX-DX}}^{'} = \alpha_{1}^{\mathrm{DD}} (V_{22}^2 V_{12} V_{32} \psi_{\mathrm{UP},+} \psi_{\mathrm{LP},+} \hat{a}_{2,+}^\dagger \hat{a}_{1,+}^\dagger + h.c.) \\ \notag
& + \alpha_{1}^{\mathrm{DD}} (V_{22}^2 V_{32}^2 \psi_{\mathrm{LP},-}^{'*} \psi_{\mathrm{LP},-} \hat{a}_{2,-}^\dagger \hat{a}_{1,-} + h.c.) \\ \notag
& + \alpha_{2}^{\mathrm{DD}} (V_{22}^2 V_{12} V_{32} \psi_{\mathrm{UP},+}\psi_{\mathrm{LP},-} \hat{a}_{2,+}^\dagger \hat{a}_{1,-}^\dagger + h.c.) \\ \notag
& + \alpha_{2}^{\mathrm{DD}} (V_{22}^2 V_{32}^2 \psi_{\mathrm{LP},-}^{*}\psi_{\mathrm{LP},-}^{'} \hat{a}_{1,+}^\dagger \hat{a}_{2,+} + h.c.) \\ \notag
& + \alpha_{2}^{\mathrm{DD}} (V_{22}^2 V_{32}^2 \psi_{\mathrm{LP},+}^{*}\psi_{\mathrm{LP},-} \hat{a}_{1,-}^\dagger \hat{a}_{1,+} + h.c.).
\end{align}

A similar procedure can be applied to IX-IX and DX-IX interactions, leading to the transformed interaction Hamiltonian of the form:
\begin{align}
\label{eq:H_int_RWA}
\hat{\mathcal{H}}_{\mathrm{int}}^{'} &= (A \psi_{\mathrm{UP},+} \psi_{\mathrm{LP},+} \hat{a}_{2,+}^\dagger \hat{a}_{1,+}^\dagger + h.c.) \\ \notag &+ (B \psi_{\mathrm{LP},-}^{'*} \psi_{\mathrm{LP},-} \hat{a}_{2,-}^\dagger \hat{a}_{1,-} + h.c.) \\ \notag &+ (C \psi_{\mathrm{UP},+}\psi_{\mathrm{LP},-} \hat{a}_{2,+}^\dagger \hat{a}_{1,-}^\dagger + h.c.) \\ \notag &+ (D \psi_{\mathrm{LP},-}^{'*}\psi_{\mathrm{LP},-} \hat{a}_{2,+}^\dagger \hat{a}_{1,+} + h.c.)\\ \notag &+ (E \psi_{\mathrm{LP},+}^{*}\psi_{\mathrm{LP},-} \hat{a}_{1,-}^\dagger \hat{a}_{1,+} + h.c.),
\end{align}
where we defined the constants
\begin{align}
\notag
&A= \alpha_1^{\mathrm{DD}} V_{22}^2 V_{12} V_{32} + \alpha_1^{\mathrm{II}} V_{23}^2 V_{13} V_{33} + \alpha_1^{\mathrm{DI}} V_{22} V_{23} V_{12} V_{33},\\ \label{eq:constants}
&B= \alpha_1^{\mathrm{DD}} V_{22}^2 V_{32}^2 + \alpha_1^{\mathrm{II}} V_{23}^2 V_{33}^2 + \alpha_1^{\mathrm{DI}} V_{22}^2 V_{33}^2,\\ \notag
&C = \alpha_2^{\mathrm{DD}} V_{22}^2 V_{12} V_{32} + \alpha_2^{\mathrm{II}} V_{23}^2 V_{13} V_{33} + \alpha_2^{\mathrm{DI}} V_{22} V_{23} V_{12} V_{32},\\ \notag
&D = \alpha_2^{\mathrm{DD}} V_{22}^2 V_{32}^2 + \alpha_2^{\mathrm{II}} V_{23}^2 V_{33}^2 + \alpha_2^{\mathrm{DI}} V_{23}^2 V_{32}^2,\\ \notag
&E = \alpha_2^{\mathrm{DD}} V_{22}^2 V_{32}^2 + \alpha_2^{\mathrm{II}} V_{23}^2 V_{33}^2 + \alpha_2^{\mathrm{DI}} V_{22} V_{23} V_{33}^2.
\end{align}

To achieve the desired interaction between quantum modes we perform a polarization transformation from the circular to linear basis, accounting for the TE-TM splitting, which is present in exciton-polaritonic systems. The transformation reads:
\begin{align}
&\hat{a}_{n,+} = \frac{1}{\sqrt{2}}\left( \hat{a}_{n,\mathrm{TM}}e^{2i\phi_n} + i \hat{a}_{n,\mathrm{TE}}e^{2i\phi_j} \right), \\ \notag
&\hat{a}_{n,-} = \frac{1}{\sqrt{2}}\left( \hat{a}_{n,\mathrm{TM}}e^{-2i\phi_n} - i \hat{a}_{n,\mathrm{TE}}e^{-2i\phi_n} \right),
\end{align}
where $\phi_n$ ($n=1,2$) represent angles encoding momentum states. We are mainly interested in the TM interaction terms (target quantum modes $\hat{a}_{1,\mathrm{TM}}$ and $\hat{a}_{2,\mathrm{TM}}$), while TE modes as well as cross-terms can be disregarded in the case of large TE-TM splitting (see the discussion in the next section). Also, we write the classically driven modes $\psi_{j}$ explicitly as complex numbers with absolute value of $\Psi_j$ and phase $\phi_j$, namely:
\begin{align}
\psi_{\mathrm{UP},+} = \Psi_{\mathrm{UP},+}e^{i \phi_{\mathrm{UP},+}},\quad \psi_{\mathrm{UP},+} = \Psi_{\mathrm{LP},+}e^{i \phi_{\mathrm{LP},+}},\\ \notag
\psi_{\mathrm{LP},-} = \Psi_{\mathrm{LP},-}e^{i \phi_{\mathrm{LP},-}},\quad \psi_{\mathrm{LP},-}^{'} = \Psi_{\mathrm{LP},-}^{'}e^{i \phi_{\mathrm{LP},-}^{'}}.
\end{align}
Performing the transformation of Hamiltonian (\ref{eq:H_int_RWA}), we get:
\begin{align}
\label{eq:H_int_fin}
&\hat{\mathcal{H}}_{\mathrm{int}}^{''} = \bigg[ \frac{A}{2} \Psi_{\mathrm{UP},+} \Psi_{\mathrm{LP},+} e^{i(\phi_{\mathrm{UP},+}+\phi_{\mathrm{LP},+}-2\phi_2 - 2\phi_1)} \\ \notag
&+ \frac{C}{2} \Psi_{\mathrm{UP},+} \Psi_{\mathrm{LP},-} e^{i(\phi_{\mathrm{UP},+}+\phi_{\mathrm{LP},-}-2\phi_2 + 2\phi_1)} \bigg] \hat{a}_{2,\mathrm{TM}}^\dagger \hat{a}_{1,\mathrm{TM}}^\dagger \\ \notag
&+ \bigg[ \frac{B}{2} \Psi_{\mathrm{LP},-}^{'} \Psi_{\mathrm{LP},-} e^{i(\phi_{\mathrm{LP},-}-\phi_{\mathrm{LP},-}^{'}+2\phi_2 - 2\phi_1)} \\ \notag
&+ \frac{D}{2} \Psi_{\mathrm{LP},-}^{'} \Psi_{\mathrm{LP},-} e^{i(\phi_{\mathrm{LP},-}-\phi_{\mathrm{LP},-}^{'}-2\phi_2 + 2\phi_1)} \bigg] \hat{a}_{2,\mathrm{TM}}^\dagger \hat{a}_{1,\mathrm{TM}} \\ \notag
&+ \frac{E}{2} \Psi_{\mathrm{LP},+} \Psi_{\mathrm{LP},-} e^{i(\phi_{\mathrm{LP},-}-\phi_{\mathrm{LP},+}+4\phi_1)} \hat{a}_{1,\mathrm{TM}}^\dagger \hat{a}_{1,\mathrm{TM}} + h.c.,
\end{align}
where the first two terms correspond to useful parametric and linear coupling terms, and the third term represents an additional energy shift of one of the modes. Finally, to reduce the system to the required $\hat{q}_1 \hat{p}_2$ type of interaction, the phases of the classical drives can be adjusted to make each term in (\ref{eq:H_int_fin}) purely imaginary. This can be satisfied with two sets of conditions defined by the right hand side of Eq. (\ref{eq:cond_4}):
\begin{align}
\label{eq:cond_1}
&e^{i(\phi_{\mathrm{UP},+}+\phi_{\mathrm{LP},+}-2\phi_2 - 2\phi_1)} = i,\\
\label{eq:cond_2}
&e^{i(\phi_{\mathrm{UP},+}+\phi_{\mathrm{LP},-}-2\phi_2 + 2\phi_1)}= i,\\
\label{eq:cond_3}
&e^{i(\phi_{\mathrm{LP},-}-\phi_{\mathrm{LP},-}^{'}+2\phi_2 - 2\phi_1)} = i,\\
\label{eq:cond_4}
&e^{i(\phi_{\mathrm{LP},-}-\phi_{\mathrm{LP},-}^{'}-2\phi_2 + 2\phi_1)} = \pm i.
\end{align}
The first system of equations (with plus sign) can be satisfied for
\begin{align}
&\phi_2 = \phi_1 + n \pi/2,\\ \notag
&\phi_{\mathrm{UP},+} = - \phi_{\mathrm{LP},-} + (2n+1) \pi/2,\\ \notag
&\phi_{\mathrm{LP},+} = \phi_{\mathrm{LP},-} + 4\phi_1,\\ \notag
&\phi_{\mathrm{LP},-}^{'} = \phi_{\mathrm{LP},-} - (2n+1)\pi/2,
\end{align}
and we have the freedom in choosing $\phi_{\mathrm{LP},-}$ phase. For this choice of phases, the interaction constant shall be tuned to $(A \Psi_{\mathrm{UP},+} \Psi_{\mathrm{LP},+} + C \Psi_{\mathrm{UP},+} \Psi_{\mathrm{LP},-})/2 = (B + D) \Psi_{\mathrm{LP},-}^{'} \Psi_{\mathrm{LP},-}/2 \equiv J$.

The second system of equations (with minus sign) can be satidfied for
\begin{align}
&\phi_2 = \phi_1 + n \pi/2 + \pi/4,\\ \notag
&\phi_{\mathrm{UP},+} = - \phi_{\mathrm{LP},-} + (n+1) \pi,\\ \notag
&\phi_{\mathrm{LP},+} = \phi_{\mathrm{LP},-} + 4\phi_1,\\ \notag
&\phi_{\mathrm{LP},-}^{'} = \phi_{\mathrm{LP},-} - n \pi/2,
\end{align}
with interaction constant tuned to $(A \Psi_{\mathrm{UP},+} \Psi_{\mathrm{LP},+} + C \Psi_{\mathrm{UP},+} \Psi_{\mathrm{LP},-})/2 = (B - D) \Psi_{\mathrm{LP},-}^{'} \Psi_{\mathrm{LP},-}/2 \equiv J$.
Accounting for the possibility to modify the system on the fly with adjustable pumps, we can in principle organize a sequence of gates between eight different momentum modes $\Phi_1 = \{ \phi_1, \phi_1 + \pi/4, ..., \phi_1 + 7\pi/4 \}$ characterized by wave vectors $\mathbf{k}_1$. Finally, coupling to another subspace of continuous wave modes $\Phi_1'$ defined by $\phi_1'$ can be done with lower fidelity and exploiting error correction afterwards.

Given the versatility of the system, we can efficiently control the couplings by classical drive amplitudes, and thus arrange the $\hat{\mathcal{H}} = 2 J \hat{q}_{1,\mathrm{TM}}\hat{p}_{2,\mathrm{TM}}$ Hamiltonian for original interaction constants $\alpha_{1,2}^{nm}$ ($nm=\mathrm{DD},\mathrm{II},\mathrm{DI}$), including both triplet and singlet interactions. Additionally, the phase conditions set the $\hat{a}_{1,\mathrm{TM}}^\dagger \hat{a}_{1,\mathrm{TM}}$ energy shift to $E\Psi_{\mathrm{LP},+}\Psi_{\mathrm{LP},-}$, implying that it shall be minimized by weak drive conditions for $\psi_{\mathrm{LP},+}$ and $\psi_{\mathrm{LP},-}$ classical modes, and stronger pumping of $\psi_{\mathrm{UP},+}$ and $\psi_{\mathrm{LP},-}^{'}$ modes.\\

\noindent {\bf 4. Multi gate realization: TE mode influence}\\

Now, let us return to the question of the rotating wave approximation (RWA) validity for the TE-TM cross-interaction terms. So far we kept only TM modes, assuming that the TE-TM interaction is strongly suppressed due to the energy shift of TE modes, i.e. TE-TM splitting. However, if the splitting $\Delta_{\mathrm{TE-TM}}$ is small comparing to other relevant energy scales (interaction constants and decay rate), this assumption becomes invalid. To test the limits in which auxiliary TE modes can be neglected, we recall that together with useful terms appearing in (\ref{eq:H_int_fin}), various spurious terms appear, namely: $\epsilon_{11}\hat{a}_{1,\mathrm{TE}}^\dagger\hat{a}_{1,\mathrm{TE}}$, $\epsilon_{22}\hat{a}_{2,\mathrm{TE}}^\dagger\hat{a}_{2,\mathrm{TE}}$, ($\epsilon_{12}\hat{a}_{1,\mathrm{TE}}^\dagger\hat{a}_{2,\mathrm{TE}}$+h.c.), ($\eta_{21} \hat{a}_{1,\mathrm{TM}}^\dagger \hat{a}_{2,\mathrm{TE}}^\dagger$+h.c), ($\eta_{12} \hat{a}_{1,\mathrm{TM}}^\dagger \hat{a}_{1,\mathrm{TE}}^\dagger$+h.c), ($\zeta_{21} \hat{a}_{2,\mathrm{TM}}^\dagger \hat{a}_{1,\mathrm{TE}}$+h.c), etc. Here constants $\epsilon_{ij}$, $\eta_{ij}$ and $\zeta_{ij}$ denote generic coupling represented by functions of bare interactions, Hopfield coefficients, and classical drive amplitudes. The first three terms act fully in the extra subspace of TE modes and are irrelevant for our considerations. However, the terms of fourth, fifth and so on type produce the parasitic rotation for the system modes $\hat{a}_{1/2,\mathrm{TM}}$, or alternatively effective leakage to the additional TE mode subspace. The related fidelity degradation of the gate then depends on the values of couplings, mode detuning $\Delta_{\mathrm{TE-TM}}$, and decay of the mode.
\begin{figure}
\includegraphics[width=0.8 \linewidth]{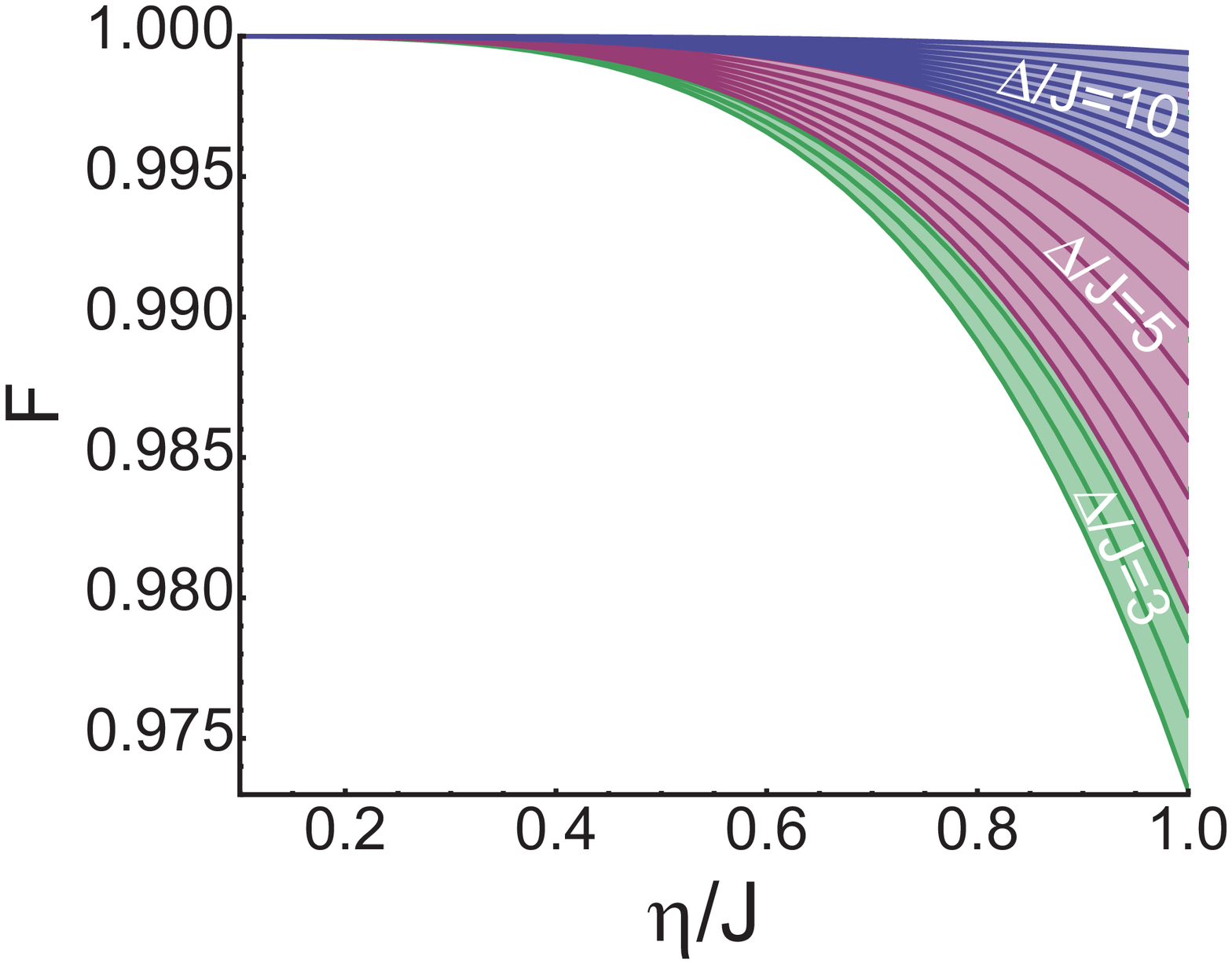}
\caption{Influence of a spurious TE-TM couplings on the fidelity of multimode CNOT gate. The minimal fidelity is shown as function of additional coupling $\eta$ for various mode occupations ($\langle x\rangle^2=1$ to $\langle x\rangle^2=10$), and three values of dimensionless TE-TM splitting.}
\label{fig:TE-TM}
\end{figure}

To quantify the fidelity change due to additional couplings, we refrain from considering a particular system with defined coupling, but characterize the generic influence of e.g. $\eta$ type of the coupling. For this, we consider Hamiltonian:
\begin{align}
\label{eq:H_TE_0}
&\hat{\mathcal{H}}=J (-i\hat{a}_{1,\mathrm{TM}}\hat{a}_{2,\mathrm{TM}}+i\hat{a}^\dagger_{1,\mathrm{TM}}\hat{a}^\dagger_{2,\mathrm{TM}}-i\hat{a}_{1,\mathrm{TM}}^\dagger \hat{a}_{2,\mathrm{TM}}\\ \notag &+i\hat{a}_{2,\mathrm{TM}}^\dagger \hat{a}_{1,\mathrm{TM}}) + \eta (\hat{a}_{2,\mathrm{TM}}^\dagger \hat{a}_{1,\mathrm{TE}}^\dagger + h.c.) + \Delta \hat{a}_{1,\mathrm{TE}}^\dagger\hat{a}_{1,\mathrm{TE}},
\end{align}
where $\Delta$ is a generic TE mode shift defined by TE-TM splitting and nonlinear contribution. Here, for the sake of simplicity let us rename modes as $\hat{a}_{1,\mathrm{TM}}\equiv \hat{a}_1$, $\hat{a}_{2,\mathrm{TM}}\equiv \hat{a}_2$, and $\hat{a}_{1,\mathrm{TE}}\equiv \hat{a}_3$. Next, Eq. (\ref{eq:H_TE_0}) can be rewritten using position and momentum operators $\hat{q}_j$ and $\hat{p}_j$ associated to each mode, which gives:
\begin{align}
\label{eq:H_TE_1}
\hat{\mathcal{H}}=&2J \hat{q}_1 \hat{p}_2 - \eta (\hat{q}_2 \hat{p}_3 + \hat{p}_2 \hat{q}_3) + \frac{\Delta}{2} (\hat{q}_3^2 + \hat{p}_3^2).
\end{align}
Deriving the equations of motion for average amplitudes $\langle \hat{q}_j \rangle$, $\langle \hat{p}_j \rangle$ ($j=1,2,3$) and associated correlators, we can calculate the fidelity for CNOT gate acting in the $\{\hat{a}_1,\hat{a}_2\}$ mode subspace as a function of dimensionless parameters $\eta/J$ and $\Delta/J$.

The results are shown in Fig. \ref{fig:TE-TM}, where we considered the cavity decay rate to be small, $\Gamma/J \ll 1$. We see that even for spurious interaction constants $\eta$ being comparable to coupling $J$, the degradation of fidelity can be suppressed by the shift of TE mode. In particular, taking $J = 0.1$ meV for a dipolaritonic system, and assuming realistic $0.5$ meV TE-TM splitting one can achieve the 0.99 fidelity commensurable with previously anticipated degradation due to decay of the cavity mode.

\end{document}